\documentclass{llncs}
\usepackage{amsmath,amssymb,bm,latexsym} % 数学記号用パッケージ
\usepackage{subfig} % subfigure用パッケージ
\usepackage{ascmac} % 文字枠用パッケージ
\usepackage{algorithm}
\usepackage{algorithmic}

\newtheorem{defi}{Definition}

\begin{document}
\title{Fourier-based Function Secret Sharing\\ with General Access Structure}
\author{Takeshi Koshiba}
\institute{Faculty of Education and Integrated Arts and Sciences,\\
Waseda University, Tokyo, Japan\\[5pt]
\tt \email tkoshiba@waseda.jp}
\maketitle

\begin{abstract}
Function secret sharing (FSS) scheme
%,formally introduced by Boyle et al.\ at EUROCRYPT 2015,
is a mechanism that calculates a function $f(x)$ 
for $x\in \{0,1\}^n$ which is shared among $p$ parties, by using 
distributed functions $f_{i}:\{0,1\}^n \rightarrow \mathbb{G}$ 
$(1\leq i\leq p)$, where $\mathbb{G}$ is an Abelian group, 
while the function $f:\{0,1\}^n \rightarrow \mathbb{G}$ 
is kept secret to the parties.
Ohsawa et al.\ in 2017 observed that any function $f$ can be described 
as a linear combination of the basis functions by regarding the function 
space as a vector space of dimension $2^n$
and gave new FSS schemes based on the Fourier basis.
All existing FSS schemes are of $(p,p)$-threshold type. That is,
to compute $f(x)$, we have to collect $f_i(x)$ for all the
distributed functions. In this paper, as in the secret sharing schemes,
we consider FSS schemes with any general access structure.
To do this, we observe that Fourier-based FSS schemes by Ohsawa et al.\
are compatible with linear secret sharing scheme. By 
incorporating the techniques of linear secret sharing with any general access
structure into the Fourier-based FSS schemes, we show Fourier-based
FSS schemes with any general access structure.
\end{abstract}


\section{Introduction}
Secret sharing (SS) schemes are fundamental cryptographic primitives,
which were independently invented 
by Blakley \cite{Blakley79} and Shamir \cite{Shamir79}.
SS schemes involve several ordinary parties (say, $p$ parties) and
the special party called a dealer. 
We suppose that the dealer has a secret information $s$
and partitions the secret information $s$ into
share information $S_{i}$ $(0\leq i \leq p)$ which will be
distributed to the $i$-th party.
In $(n,p)$-threshold SS scheme, the secret information $S$ can be
recovered from $n$ shares (collected if any $n$ parties get together),
but no information on $s$ is obtained
from at most $n-1$ shares. This threshold property can be discussed
in terms of access structures. An access structure $({\cal A},{\cal B})$ 
consists of two classes of sets of parties such that (1) if all parties
in some set $A\in {\cal A}$ get together then the secret information
can be recovered from their shares; (2) even if all parties
in any set $B\in {\cal B}$ get together then any information of the secret $s$
cannot be obtained. For example, the access structure $({\cal A},{\cal B})$ 
of the $(n,p)$-threshold SS scheme can be defined as
${\cal A} = \{ A\subseteq \{1,\ldots, p\}  : |A| \ge n \}$ 
and ${\cal B} = \{ B\subseteq \{1,\ldots, p\} : |B| < n \}$.
Besides the access structure of the threshoold type, many variants have
been investigated in the literature \cite{BL88,Brickell89e,Brickell89c,ISN87,Kothari84,NNP03}. As a standard technique for constructing access structures, 
{\em monotone span programs} \cite{KW93,Fehr98,Fehr99,NN04} are often used.

% PIRの例は難しいか？何か例をあげて，FSSの動機を説明
The idea where a secret information is secretly disributed to several parties
can be applied to a function.
The idea of secretly distributing a function has 
an application in private information retrieval (PIR) \cite{CG97,CGKS98,KO97}
as demonstrated in \cite{b}.
%For example, we can know how often some keyword $w$ appears 
%in a database $D$ without revealing the keyword $w$.
%A client first distributes the point function $f_{w,1}$ into $k$ 
%distributed functions $(f_1, \ldots , f_k)$.
%Each distributed database calculates the share value 
%$f_i(w)$ and sends it back to the client.
%After that, the client can obtain $f_{w,1}(w)$ from all shares.
%If an FSS scheme can distribute arbitrary functions,
%it enables us to compute the functional value 
%without revealing any information on the algorithm to evaluate
%the function to servers.
%what giving measure of algorithm by distributing algorithm functions 
%and administering plurality of servers.
%
%FSS の歴史
%流れとしてはGilboa -> Boyle -> Ohsawa その後 computational な設定の話
Gilboa et al.~\cite{b} consider to distribute point functions 
(DPFs) $f_{a,b}:\{0,1\}^n \rightarrow \mathbb{G}$,
where $f_{a,b}(x)=b$ if $x=a$ for some $a \in \{0, 1\}^n$ and $f_{a,b}(x)=0$ 
otherwise.
In a basic DPF scheme, the function $f$ is partitioned into two keys 
$f_{0}, f_{1}$ and each key is distributed to the respective party of 
the two parties. Each party calculates the share 
$y_{i}=f_i(x)$ for common input $x$ by using the key $f_i$.
On the other hand, each $f_i$ does not give any important information
(e.g., the value $a$ for $f_{a,b}$) on the original function.
The functional value of the point function $f_{a,b}$
can be obtained by just summing up two shares $y_0$ and $y_1$ of 
the two parties.
Boyle et al.~\cite{a} investigate the efficiency in the key size and 
extend the two-party setting into the multi-party setting. 
Moreover, they generalize 
the target functions (i.e., point functions) to other functions,
and propose an FSS scheme for some function family $\mathcal{F}$ 
in which functions $f:\{0,1\}^n \rightarrow \mathbb{G}$ can be calculated 
efficiently. 
In the multi-key FSS scheme we partition
a function $f \in \mathcal{F}$ into $p$ distributed functions 
$(f_{1}, \ldots, f_{p})$. Likewise,
an equation $f(x)=\sum ^p_{i=1}f_{i}(x)$ is satisfied with respect to any $x$,
and the information about the secret function $f$ (except the domain and 
the range) does not leak out from at most $p-1$ distributed functions.
Moreover, distributed functions $f_{i}$ can be described as 
short keys $k_{i}$ and it is required to be efficiently evaluated.

In \cite{OKK17}, Ohsawa et al. observed that any function $f$
from $\{0,1\}^n$ to $\{0,1\}$ can be described as a linear combination
of the basis functions by regarding the function space as a vector space
of dimension $2^n$. While the point functions $f_{a,1}$ 
(for all $a\in \{0,1\}^n$) constitute a (standard) basis for the vector space,
any function $f:\{0,1\}^n\rightarrow \{\pm 1\}$ can be represented
as a linear combination of the Fourier basis functions 
$\chi_a(x)=(-1)^{\langle a,x\rangle}$, where $\langle a, x\rangle$
denotes the inner product between vectors $a=(a_1,\ldots, a_n)$ and 
$x=(x_1,\ldots, x_n)$. Based on the above observation,
Ohsawa et al.\  gave new FSS schemes based on the Fourier basis.
If we limit our concern to polynomial-time computable FSS schemes, 
functions for which the existing schemes are available would be limited.
Since polynomial-time computable functions represented 
by combinations of point functions 
are quite different from ones represented by the Fourier basis functions,
point function based FSS schemes and Fourier function based FSS schemes
are complementary.

We note that properties of some functions are often
discussed in the technique of the Fourier analysis.
Akavia, Goldwasser and Safra~\cite{AGS03} introduced a
novel framework for proving hard-core properties
in terms of Fourier analysis. Any predicates can be represented
as a linear combination of Fourier basis functions. Akavia et al.\
show that if the number of non-zero coefficients
in the Fourier representation of hard-core predicates
is polynomially bounded then the coefficients are 
efficiently approximable. This fact leads to the hard-core properties.
Besides hard-core predicates, it is well known that 
low-degree polynomials are Fourier-concentrated \cite{Obook}.

%Boyle et al.\ constructs FSS schemes for interval functions and for
%partial matching functions by the essential use of pseudo-random generators 
%(PRG) like the Goldreich-Goldwasser-Micali \cite{GGM86} 
%construction of pseudo random functions. They also show that FSS schemes 
%for functions which can be calculated in polynomial time 
%are implemented by using some obfuscation techniques and one-way functions.
%Furthermore, Boyle et al.~\cite{BGI16} recently construct a two-party
%FSS scheme for {\it branching programs} under the DDH assumption, which
%implies low-communication two-party secure computation protocols.
%A $p$-party FSS scheme for {\it circuits} was proposed in
%\cite{DHR16} under the LWE assumption, by using
%multi-key fully-homomorphic encryption.

% ここは上のパラグラフと下のパラグラフを上手く融合して，
% かつ，our distribution の箇所も纏めて，歴史を構成する。
\subsection*{Contribution}
Since the existing FSS schemes are of $(p,p)$-threshold type, it is
natural to consider the possibility of FSS schemes with 
any threshold structure of $(n,p)$-type and even general
access structures as in the case of SS schemes.

In this paper, we affirmatively answer to this question. 
As mentioned, Fourier-based FSS schemes in \cite{OKK17} are quite simpler
than the previous FSS schemes. This is because Fourier basis functions
have some linear structure. Shamir's threshold SS scheme can be seen
as an application of the Reed-Solomon code, which is a linear code.
Both the distribution phase and the reconstruction phase can be described
in a linear algebraic way. From this viewpoint, we construct an
$(n,p)$-threshold Fourier-based FSS scheme. Moreover, SS schemes
with general access structure can be discussed in terms of
monotone span program (MSP). The underlying structure of
SS schemes by using MSP is similar to the linear algebraic view
of Shamir's $(n,p)$-threshold SS scheme, we can similarly
construct Fourier-based FSS schemes with general access structure.

Technically speaking, Ohsawa et al.\cite{OKK17} consider a function
from $\{0,1\}^n$ to $\mathbb{C}$. That is, they consider 
Fourier transform over $n$-dimensional vector space of $\mathbb{F}_2$.
On the other hand, we consider a function from 
a finite field $\mathbb{F}_q$ (of prime order $q$) to $\mathbb{C}$. So,
in this paper, we consider the Fourier transform over $\mathbb{F}_q$
rather than $(\mathbb{F}_2)^n$.
The shift of the underlying mathematical structure enables to
construct FSS schemes with general access structure.

\section{Preliminaries}
\subsection{Access Structure and Monotone Span Program}
Let us assume that there are $p$ parties in an SS (or, FSS) scheme.
A {\em qualified} group is a set of parties
who are allowed to reconstruct the secret and a {\em forbidden}
group is a set of parties who should not be able to get
any information about the secret. The set of qualified groups
is denoted by $\cal A$ and the set of forbidden groups by $\cal B$.
The set $\cal A$ is said to be {\em monotonically increasing} if,
for any set $A\in \cal A$, any set $A'$ such that $A'\supseteq A$
is also included in $\cal A$. The set $\cal B$ is said to be
{\em monotonically decreasing} if, for any set $B\in\cal B$,
any set $B'$ such that $B'\subseteq B$ is also included in $\cal B$.
If a pair $({\cal A}, {\cal B})$ satisfies that ${\cal A}\cap {\cal B}
= \varnothing$, $\cal A$ is monotonically increasing and $\cal B$ is
monotonically decreasing, then the pair is called a
(monotone) {\em access structure}. If an access structure $({\cal A},{\cal B})$
satisfies that ${\cal A} \cup {\cal B}$ coincides with the power set of
$\{1,\ldots, p\}$, we say that the access structure is {\em complete}.
If we consider a complete access structure, we may simply denote 
the access structure by $\cal A$ instead of $({\cal A}, {\cal B})$,
since $\cal B$ is equal to the complement set of $\cal A$.

As mentioned, there are several ways to realize general access structures.
Monotone span program (MSP) 
is a typical way to construct general access structures.
Before mentioning the MSP, we prepare some basics and notations for
linear algebra.

An $m\times d$ matrix $M$ over a field $\mathbb{F}$ defines a linear map
from $\mathbb{F}^d$ to $\mathbb{F}^m$. The {\em kernel} of $M$, denoted
by ${\rm ker}(M)$, is the set of vectors $\boldsymbol{u}\in\mathbb{F}^d$ 
such that $M\boldsymbol{u}=\boldsymbol{0}$. 
The {\em image} of $M$, denoted by ${\rm im}(M)$, is the set of
vectors $\boldsymbol{v}\in\mathbb{F}^m$ such that
$\boldsymbol{v}=M\boldsymbol{u}$ for some $\boldsymbol{u}\in\mathbb{F}^d$.

A monotone span program (MSP) $\cal M$ is a triple
$(\mathbb{F}, M, \rho)$, where $\mathbb{F}$ is a finite field, $M$ is an
$m\times d$ matrix over $\mathbb{F}$,
$\rho: \{1,\ldots,m\}\rightarrow \{1,\ldots, p\}$ is a surjective
function which labels each row of $M$ by a party.
For any set $A\subseteq \{1,\ldots, p\}$, let $M_A$ denote
the sub-matrix obtained by restricting $M$ to the rows labeled by parties
in $A$. We say that $\cal M$ accepts $A$ 
if $\boldsymbol{e}_1=(1,0,\ldots, 0)^T\in {\rm im}(M_A^T)$, 
otherwise we say $\cal M$ rejects $A$.
Moreover, we say that $\cal M$ accepts a (complete) access structure $\cal A$
if the following is equivalent: $\cal M$ accepts $A$ if and only if
$A\in \cal A$.

When $\cal M$ accepts a set $A$, there exists a {\em recombination}
vector $\boldsymbol{\lambda}$ such that 
$M_A^T\boldsymbol{\lambda}=\boldsymbol{e}_1$.
Also note that $\boldsymbol{e}_1\not\in {\rm im}(M_B^T)$ if and only if
there exists a vector $\boldsymbol{\xi}$ such that
$M_B\boldsymbol{\xi}=\boldsymbol{0}$ and the first element
of $\boldsymbol{\xi}$ is 1.

\subsection{Function Secret Sharing}
The original definition in \cite{a} of FSS schemes
are tailored for threshold schemes. We adapt the definition
for general access structures.
In an FSS scheme, we partition a function $f$ into keys $k_i$ 
(the succinct descriptions of $f_i$)
which the corresponding parties $P_i$ receive.
Each party $P_i$ calculates the share $y_i=f_i(x)$
for the common input $x$.
The functional value $f(x)$ is recovered from shares 
$\vec{y}_A$ in a qualified set $A$ of parties, which is a 
sub-vector of $\vec{y}= (y_1, y_2, \ldots , y_p)$,
by using a decode function {\it Dec}.
Any joint keys $k_i$ in a forbidden set $B$ of parties
do not leak any information on function $f$
except the domain and the range of $f$.
We first define the decoding process from shares.
\begin{defi}\rm
 (Output Decoder)
An output decoder ${\it Dec}$, on input
a set $T$ of parties and shares from the parties in $T$,
outputs a value in the range $R$ of the target function $f$.
\end{defi}
Next, we define FSS schemes.
We assume that $\cal A$ is a complete access structure
among $p$ parties and $T\subseteq \{1,2,\ldots, p\}$ be a set of parties.
\begin{defi}\rm
For any $p \in \mathbb{N}$, $T \subseteq \{1,2,\ldots, p\}$, 
an $\cal A$-secure FSS scheme with respect to a
function class $\cal F$ is a pair of 
PPT algorithms $({\it Gen}, {\it Eval})$ satisfying the following.
 \begin{itemize}
  \item The key generation algorithm ${\it Gen}(1^\lambda ,f)$, 
on input the security parameter
 $1^\lambda$ and a function $f:D\rightarrow R$ in $\cal F$, 
% the key generation algorithm $\it Gen$
outputs $p$ keys $(k_1,\ldots, k_p)$.
       \item 
The evaluation algorithm
${\it Eval}(i,k_i,x)$, on input a party index $i$,
a key $k_i$, and an element $x \in D$, 
%the evaluation algorithm $\it Eval$ 
outputs 
a value $y_i$, corresponding to the $i$-th party's share of $f(x)$.
\end{itemize}
Moreover, these algorithms must satisfy the following properties:
%{\em Correctness} and {\em Security}.
\begin{itemize}
 \item ${\it Correctness}$: For all $A\in \cal A$, 
$f \in  \mathcal{F}$ and $x \in D$,
       \begin{eqnarray*}
\Pr[
{\it Dec}(A, \{ {\it Eval}(i, k_i, x) \}_{i\in A})=f(x)
\mid
(k_1,\ldots, k_p)\leftarrow {\it Gen}(1^\lambda, f)]=1.
       \end{eqnarray*}
 \item ${\it Security:}$
       Consider the following indistinguishability challenge experiment 
for a forbidden set $B$ of parties, where $B\not\in \cal A$:
       \begin{enumerate}
	\item The adversary $\cal D$ outputs $(f_0, f_1)
\leftarrow\mathcal{D}(1^\lambda)$,
where $f_0, f_1\in \cal F$.
	\item The challenger chooses $b \leftarrow \{0,1\}$
	and $(k_1, \ldots, k_p) \leftarrow {\it Gen}(1^\lambda, f_b)$.
	      \item $\mathcal{D}$ outputs a guess 
	$b' \leftarrow \mathcal{D}(\{k_i\}_{i\in B})$, 
	given the keys for the parties in the forbidden set $B$.
       \end{enumerate}       
The advantage of the adversary $\cal D$ is defined as 
${\it Adv}(1^\lambda, \mathcal{D}):= \Pr[b=b']- 1/2$.
The scheme $({\it Gen}, {\it Eval})$ satisfies that
there exists a negligible function $\nu$ such that for all non-uniform 
PPT adversaries $\mathcal{D}$ which corrupts parties in any forbidden
set $B$, it holds that 
${\it Adv}(1^\lambda, \mathcal{D})\leq \nu (\lambda)$.
\end{itemize}
\end{defi}

\subsection{Basis functions}\label{sec:basis_function}
The function space of functions $f:\mathbb{F}_q \rightarrow \mathbb{C}$ 
can be regarded as a vector space of dimension $q$.
Therefore, the basis vectors for the function space exist 
and we let $h_{i}(x)$ be each basis function.
Any function $f$ in the function space is described as a 
linear combination of the basis functions
\begin{eqnarray*}
f(x) = \sum_{j\in \mathbb{F}_q} \beta_j h_j(x),
\end{eqnarray*}
where $\beta_j$'s are coefficients in $\mathbb{C}$.
%
%The point functions mentioned above can be considered as basis
%functions, so we can describe any functions by a linear combination 
%of the point functions.

\subsubsection*{The Fourier basis}~\\
Let $f:\mathbb{F}_q \rightarrow \mathbb{C}$, where $q$ is an odd prime number.
The Fourier transform of the function $f$ is defined as
\begin{eqnarray}\label{sec:1}
\hat{f}(a)=\frac{1}{q}\sum_{x\in \mathbb{F}_q} f(x)e^{-2 \pi (ax/q) i},
\end{eqnarray}
where $i$ is the imaginary number.
Then, $f(x)$ can be described as a linear combination
of the basis functions $\chi_a(x)=e^{2\pi (ax/q) i}$, that is,
\[
f(x)=\sum_{a\in\mathbb{F}_q} \hat{f}(a)\chi_a(x). 
\]
In the above, $\hat{f}(a)$ is called Fourier coefficient of $\chi_a(x)$.
%As a consequence of Euler's formula $e^{\pi i} = -1$ we have
By using $\omega_q=e^{(2\pi/q) i}$, the primitive root of unity of order $q$,
we can denote each Fourier basis function by
\begin{eqnarray*}
 \chi_{a}(x) = (\omega_q)^{ax}
\end{eqnarray*}
and let ${\cal B}_F=\{ \chi_a \mid a\in\mathbb{F}_q\}$
be the sets of all the Fourier basis functions.

It is easy to see that the Fourier basis is orthonormal 
since 
\begin{eqnarray}
 \frac{1}{q}\sum_{x\in \mathbb{F}_q}\chi_{a}(x)\chi_{b}(x)
     =\begin{cases}
       1 \ \ \ \mbox{\rm if~} a=b,\\
       0 \ \ \ \mbox{\rm otherwise}.
      \end{cases}
\end{eqnarray}
In this paper, we consider only Boolean-valued functions
and assume that
the range of the boolean function is $\{\pm 1\}$ instead of $\{0,1\}$
without loss of generality. 
That is, we regard boolean functions as
mappings from $\mathbb{F}_q$ to $\{\pm 1\}$.
Also, we have
\[ \chi_{a+b}(x)=\chi_a(x)\chi_b(x). \]
This multiplicative property plays an important role in this paper.

%Let ${\cal B}_P=\{ P_{a,1} \mid a\in\{0,1\}^n\}$
%the basis functions with respect to point functions and 

\section{Linear Secret Sharing}

\subsection{Shamir's Threshold Secret Sharing}
First, we give a traditional description of Shamir's
$(n,p)$-threshold SS scheme \cite{Shamir79}, where $p\ge n\ge 2$.
Let $s$ be a secret integer which a dealer $D$ has.
First, the dealer $D$ chooses a prime number $q>s$
and a polynomial $g(X) \in \mathbb{F}_q[X]$ of degree $n-1$.
Then, the dealer $D$ computes $s_i=(i,g(i))$ as a share for the $i$-th
party $P_i$ and sends $s_i$ to each $P_i$. 
For the reconstruction, $n$ parties get together and 
recover the secret $s$ by the Lagrange interpolation from their shares.

The above procedure can be equivalently described as follows.
Let $M$ be an $n\times p$ Vandermonde matrix and $\boldsymbol{m}_i$
be the $i$-th row in $M$. That is, 
$\boldsymbol{m}_i=(1,i,i^2,\ldots, i^{n-1})$.
Let $\boldsymbol{b}=(b_0,b_1,\ldots, b_{n-1})^T$ be 
an $n$-dimensional vector such that
$b_0=s$ and $b_1,\ldots, b_{n-1}$ are randomly chosen
elements in $\mathbb{F}_q$. Let $\boldsymbol{y}=
(s_1,s_2,\ldots, s_p)^T=M\boldsymbol{b}$.
The share $s_i$ for $P_i$ is the $i$-th element of $\boldsymbol{y}$,
that is, $s_i = \langle \boldsymbol{m}_i^T,\boldsymbol{b}\rangle$,
where $\langle\cdot,\cdot\rangle$ denotes the inner product.
Let $A$ be a subset of $\{1,2,\ldots,p\}$ which corresponds to a set
of parties. Let $M_A$ be a submatrix of $M$ obtained by collecting 
rows $\boldsymbol{m}_j$ for all $j\in A$. We similarly define 
a subvector $\boldsymbol{y}_A$ by collecting elemetnts $s_j$ 
for all $j\in A$.
Let $\boldsymbol{e}_1 = (1,0,0,\ldots,0)^T \in (\mathbb{F}_q)^n$. Then
we can uniquely determine $\boldsymbol{\lambda}$ 
such that $M_A^T\boldsymbol{\lambda} = \boldsymbol{e}_1$
by solving an equation system if and only if $|A|\ge n$.
Then, we have
\[ s = \langle \boldsymbol{b}, \boldsymbol{e}_1\rangle 
     = \langle \boldsymbol{b}, M_A^T\boldsymbol{\lambda}\rangle 
     = \langle M_A\boldsymbol{b}, \boldsymbol{\lambda}\rangle 
     = \langle \boldsymbol{y}_A, \boldsymbol{\lambda}\rangle.
\]
Since $\boldsymbol{y}_A$ corresponds to all shares for $P_j$ ($j\in A$),
we can reconstruct the secret $s$ by computing the inner product
$\langle \boldsymbol{y}_A, \boldsymbol{\lambda}\rangle$.

\subsection{Monotone Span Program and Secret Sharing}\label{sec:msplss}
Here, we give a construction of linear secret sharing (LSS)
based on Monotone Span Program (MSP). Here,
we do not mention how to construct MSP. For the construction
of MSP, see the literature, e.g., \cite{KW93,Brickell89e,Fehr98,Fehr99}.
In this paper, we will use the LSS schemes.
Since the LSS schemes imply MSPs \cite{BC94,vD94}, it is sufficient
to consider MSP-based SS schemes.

Let $s\in \mathbb{F}_q$ be a secret which the dealer $D$ has
and ${\cal M}=(\mathbb{F}_q,M,\rho)$ be 
an MSP which corresponds to a complete access structure $\cal A$.
The dealer $D$ considers to partition $s$ into several shares.
In the sharing phase, the dealer $D$ chooses 
a random vector $\boldsymbol{r}\in(\mathbb{F}_q)^{p-1}$ and 
sends a share $\langle \boldsymbol{m}_i^T, (s,\boldsymbol{r})^T\rangle$ 
to the $i$-th party. In the reconstruction phase,
using the recombination vector $\boldsymbol{\lambda}$,
any qualified set $A\in \cal A$ of parties can
reconstruct the secret as follows:
\[ \langle \boldsymbol{\lambda}, M_A(s,\boldsymbol{r})^T  \rangle = 
   \langle M_A^T \boldsymbol{\lambda}, (s,\boldsymbol{r})^T \rangle =
   \langle \boldsymbol{e}_1, (s,\boldsymbol{r})^T\rangle = s. \]
Regarding the privacy, let $B$ be a forbidden set of parties,
and consider the joint information held by the parties in $B$.
That is, $M_B\vec{b} = \vec{y}_B$, where $\vec{b}=(s,\vec{r})^T$. 
Let $s'\in \mathbb{F}_q$ be an arbitrary value and let $\vec{\xi}$ be 
a vector such that
$M_B\vec{\xi}=\vec{0}$ and the first element in $\vec{\xi}$ is equal to 1.
Then $\vec{y}_B=M_B(\vec{b}+\vec{\xi}(s'-s))$, where the first
coordinate of the vector $\vec{b}+\vec{\xi}(s'-s)$ is now equal to $s'$. 
This means that, from the viewpoint of the parties in $B$, 
their shares $\vec{y}_B$ are
equally likely consistent with any secret $s'\in \mathbb{F}_q$.

\section{Our Proposal}
%FSS schemes for some function classes 
%such as point functions, interval function, and partial pattern matching 
%functions are proposed by Boyle et al.
As mentioned, any function can be described as a linear combination of
basis functions.
If the function is described as a linear combination
of a super-polynomial number of basis functions, then
the computational cost for evaluating the function might be
inefficient. We say that a function has a {\em succinct} description
(with respect to the basis $\cal B$)
if the function $f$ is described as
%\[ 
$f(x)=\sum_{h\in {\cal B}'} \beta_h h(x)$
%\]
for some ${\cal B}'\subset {\cal B}$ such that $|{\cal B}'|$
is polynomially bounded in the security parameter.
%as a linear combination of
%a polynomial number of basis functions.
If we can find a good basis set $\cal B$, some functions may have
a succinct description with respect to $\cal B$.
We consider to take the Fourier basis as such a good basis candidate.

We will provide an FSS scheme for some function class
whose elements are functions with succinct description
with respect to the Fourier basis ${\cal B}_F$.
Since the Fourier basis has nice properties,
our FSS scheme with general access structure can be realized.

In what follow, we assume that the underlying basis is always 
the Fourier basis ${\cal B}_F$. Moreover, we assume that
${\cal M}=(\mathbb{F}_q,M,\rho)$ is an MSP which corresponds 
to a general complete access structure $\cal A$.
We will consider Fourier-based FSS schemes with this access
structure.

%than the previous FSS schemes.
%for the point function which needs binary tree 
%and can realize more practically.

%As mentioned in Section 1, Akavia et al.~\cite{AGS03} 
%show that many hard-core predicates have succinct description
%with respect to the Fourier basis ${\cal B}_F$. Thus,
%our proposal can be used as an FSS scheme for hard-core predicates.

\subsection{FSS scheme for the Fourier basis}
In this subsection, we consider to partition each
Fourier basis function $\chi_a (x) = (\omega_q)^{ax}$
into several keys. That is, we give an FSS scheme
with general access structure with respect to the function class 
${\cal B}_F$.

Our FSS scheme with respect to ${\cal B}_F$
consists of three algorithms
${\it Gen}_1^F$ (Algorithm 1),  ${\it Eval}_1^F$ (Algorithm 2), 
and ${\it Dec}_1^F$ (Algorithm 3).
${\it Gen}_1^F$ is an algorithm that divides the secret $a$ 
(for $\chi_a (x)$) into $p$ keys $(k_1, \ldots, k_p)$
as in the SS scheme with the same access structure.
Each key $k_i$ is distributed to the $i$-th party $P_i$.
%two random strings $a_0$ and $a_1$ so as to satisfy that
%that ee vectors $\vec{a},\vec{a_0},\vec{a_1}$ is satisfied equation 
Note that the secret $a$ can be recovered
from the keys $k_i$ for all $i$ in a qualified set $A\in {\cal A}$.

In ${\it Eva}l^F_1$, each party obtains the share by feeding $x$ 
to the function distributed as the key. 
${\it Dec}^F_1$ is invoked in order to obtain the Fourier basis function 
$\chi_{a}(x)$ from the shares. 

The correctness follows from
\begin{eqnarray*}
\chi_a(x) & = & (\omega_q)^{ax}\\
&=& (\omega_q)^{\langle \vec{y}_A, \vec{\lambda}\rangle x}\\
&=& (\omega_q)^{(\sum k_i\lambda_i ) x}\\
&=& \prod \left( (\omega_q)^{k_ix} \right) ^{\lambda_i}.
\end{eqnarray*}

%\noindent
%{\it Correctness}:\\
%Due to the construction, the correctness
%follows from Eq.(\ref{eq:fd}).
%
%\medskip
For the security,
we assume that an adversary $\cal D$ chooses
$(f_0, f_1)$ where $f_0=\chi_a$ and $f_1=\chi_b$.
Then the challenger chooses a random bit $c$ to select $f_c$
and invokes ${\it Gen}_1^{F}(1^\lambda, a)$ if $c=0$
and ${\it Gen}_1^{F}(1^\lambda, b)$ if $c=1$.
If $c=0$ then $a$ is divided into $p$ keys.
If $c=1$ then $b$ is divided into different $p$ keys.
From the argument in Section \ref{sec:msplss},
the guess for the secret information $a$ (resp., $b$) is a perfectly 
random guess. That is, the inputs to the adversary $\cal D$ are the
same in the two cases.
Thus, the adversary $\cal D$ cannot decide if
the target function is either $\chi_{a}(x)$ or $\chi_{b}(x)$.
It implies that
only $\cal D$ can do for guessing the random bit $c$ selected by
the challenger is just a random guess. So, $Adv(1^\lambda,{\cal D})=0$.
This concludes the security proof.

\begin{figure*}[ttt!]
%\begin{tabular}[t]{c@{\hskip 1.5em}c}
%\begin{minipage}[t]{58mm}
     \begin{algorithm}[H]
    \caption{${\it Gen}_1^{F}(1^\lambda, a)$}
   \begin{algorithmic}
    \STATE Choose a random vector $\boldsymbol{r}\in (\mathbb{F}_q)^{p-1}$
uniformly ;\\
%    \ \ \ \ $(a_{0,\ 1},\ a_{0,\ 2},\ \ldots,\ a_{0,\ n}),\ \forall i,\ j,\ a_{0,\ j}\in\{0,\ 1\}$
%    \STATE Let binary representation $\vec{a}=a_1,\ a_2,\ \cdots,\ a_n$
    \FOR{$i=1$ to $p$}
    \STATE $\boldsymbol{m}_i \leftarrow$ the $i$-th row of $M$ ;
    \STATE $k_i \leftarrow \langle \boldsymbol{m}_i, (a,\boldsymbol{r})^T \rangle$
    \ENDFOR
%{0,\ 1}\|a_{0,\ 2}||\cdots||a_{0,\ n}$
%{1,\ 1}||a_{1,\ 2}||\cdots||a_{1,\ n}$
    \STATE Return $(k_1,\ldots, k_p)$.
   \end{algorithmic}
   \end{algorithm}
%\end{minipage}
\vspace{-1cm}
%\begin{tabular}[t]{ll}
%\begin{minipage}[t]{58mm}
   \begin{algorithm}[H]
    \caption{${\it Eval}_{1}^{F}(i, k_i, x)$}
    \begin{algorithmic}
%     \STATE Let binary representation $\vec{x}=x_1,\ x_2,\ \cdots,\ x_n$
%     \STATE Parse $k_i$ as $k_i=(a_{i, 1}\|a_{i, 2}\|\dots\|a_{i,n})$
     \STATE $v_i \leftarrow (\omega_q)^{k_ix}$ ;
     \STATE Return $(i,v_i)$.
    \end{algorithmic}
   \end{algorithm}
%\end{minipage}
%\\[-1.5em]
\vspace{-1cm}
%\begin{minipage}[t]{58mm}
   \begin{algorithm}[H]
    \caption{${\it Dec}_{1}^{F}(A, \{(i,v_i)\}_{i\in A})$}
    \begin{algorithmic}
     \STATE Compute a recombination vector $\vec{\lambda}=
(\lambda_1,\ldots,\lambda_p)^T$ from $A$ ;
%     \IF {$ans=1$}
%     \STATE $ans=-1$.
%     \ELSE
%     \STATE $ans=1$.
%     \ENDIF
     \STATE Return $w=\prod_{i\in A} (v_i)^{\lambda_i}$.
    \end{algorithmic}
   \end{algorithm}
%\end{minipage}
%\end{tabular}
%\end{tabular}
\end{figure*}

\subsection{General FSS Scheme for Succinct Functions}\label{sec:basis_for_DF}
Since we do not know how to evaluate any function efficiently,
we limit ourselves to succinct functions with respect
to the Fourier basis ${\cal B}_F$.
Note that succinct functions with respect to ${\cal B}_F$
do not coincide with succinct functions with respect to
point functions. Simple periodic functions
are typical examples of succinct functions with respect to ${\cal B}_F$,
which might not be succinct functions with respect to point functions.
As mentioned, some hard-core predicates of one-way functions 
are succinct functions with respect to ${\cal B}_F$.

Let ${\cal F}_{{\cal B}_F,\ell}$ be a class of functions $f$ which
can be represented as a linear combination of $\ell$ basis
functions (with respect to ${\cal B}_F$) at most, where $\ell$ is a
polynomial in the security parameter. That is, $f$ has the following form:
\[ f(x) = \sum_{i=1}^\ell \beta_i \chi_{a_i}(x). \]
We construct an FSS scheme with general access structure
$({\it Gen}^F_{\le \ell}, {\it Eval}^F_{\le \ell}, {\it Dec}^F_{\le \ell})$ 
for a function $f\in {\cal F}_{{\cal B}_F,\ell}$ as follows.
Note that the construction is a simple adaptation of the 
Fourier-based FSS scheme over $(\mathbb{F}_2)^n$ in \cite{OKK17}. 
\begin{itemize}
\item 
${\it Gen}_{\le\ell}^{F}(1^\lambda, f):$
On input the security parameter $1^\lambda$ and a function $f$, 
the key generation algorithm (Algorithm 4) 
outputs $p$ keys $(k_1,\ldots ,k_{p})$.
\item 
${\it Eval}_{\le\ell}^{F}(i, k_i, x):$
On input a party index $i$, a key $k_{i}$, and 
an input string $x \in \mathbb{F}_q$, 
the evaluation algorithm (Algorithm 5) outputs a value $y_{i}$,
corresponding to the $i$-th party's share of $f(x)$.
\item 
${\it Dec}_{\le\ell}^{F}(A, \{ y_i\}_{i\in A}):$
On input shares $\{ y_i\}_{i\in A}$ of parties in a (possibly)
qualified set $A$, the decryption algorithm (Algorithm 6)
outputs a solution $f(x)$ for $x$.
\end{itemize}

In the above FSS scheme 
$({\it Gen}^F_{\le\ell}, {\it Eval}^F_{\le\ell}, {\it Dec}^F_{\le\ell})$
for succinct functions $f\in {\cal F}_{{\cal B},\ell}$,
we invoke FSS scheme $({\it Gen}^F_1, {\it Eval}^F_1, {\it Dec}^F_1)$ 
for basis functions ${\cal B}_F$, 
since $f$ can be represented as a linear combination of at most $\ell$
basis functions.
In this construction,
we distribute each basis function $\chi_{a_i}(x)$ and each 
coefficient $\beta_i$ as follows.
We invoke $({\it Gen}^F_1, {\it Eval}^F_1, {\it Dec}^F_1)$ 
to distribute each basis function $\chi_{a_i}(x)$
and use any SS scheme with the same access structure
to distribute each coefficient $\beta_i$.

\begin{figure*}[ttt!]
%\begin{tabular}[t]{c@{\hskip 1.5em}c}
%\begin{minipage}[t]{58mm}
 \begin{algorithm}[H]
    \caption{${\it Gen}^F_{\le\ell}(1^\lambda, f(\cdot)=\sum_{i=1}^{\ell}
\beta_i \chi_{a_i}(\cdot) )$}
    \begin{algorithmic}
     \FOR{$i=1$ to $\ell$}
     \STATE $(k^i_1, k^i_2,\ldots, k^i_p)
\leftarrow$${\it Gen}_{1}^F(1^\lambda, a_i)$ ;
     \STATE $(s^i_1, s^i_2,\ldots, s^i_p) 
\leftarrow$iThe sharing phase of some SS scheme, given $\beta_i$ ;
     \ENDFOR
     \FOR{$j=1$ to $p$}
     \STATE Set $\vec{k}_j\leftarrow(k_j^1, k_j^2,\ldots, k^{\ell}_j)$ ;
     \STATE Set $\vec{s}_j\leftarrow(s_j^1, s_j^2,\ldots, s^{\ell}_j)$ ;
     \ENDFOR
     \STATE Return $((\vec{k}_1,\vec{s}_1),\ldots,(\vec{k}_p,\vec{s}_p))$.
    \end{algorithmic}
   \end{algorithm}
%\end{minipage}
\vspace{-1cm}
%\begin{tabular}[t]{ll}
%\begin{minipage}[t]{58mm}
   \begin{algorithm}[H]
    \caption{${\it Eval}^F_{\le\ell}(i,(\vec{k}_i,\vec{s}_i), x)$}
    \begin{algorithmic}
     \FOR{$j=1$ to $\ell$}
     \STATE $y_j^i\leftarrow$${\it Eval}_{1}^F(i, k_j^i, x)$ ;
     \ENDFOR
     \STATE Set $\vec{y_i}=(y_1^i, y_2^i,\ldots, y_{\ell}^i)$ ;
     \STATE Return $(i,\vec{y}_i,\vec{s}_i)$.
    \end{algorithmic}
   \end{algorithm}
%\end{minipage}
%\\[-1.5em]
\vspace{-1cm}
%\begin{minipage}[t]{58mm}
  \begin{algorithm}[H]
    \caption{${\it Dec}^F_{\le\ell}(A, \{(i,\vec{y}_i,\vec{s}_i)\}_{i\in A})$}
    \begin{algorithmic}
     \FOR{$i=1$ to $\ell$}
     \STATE $g_i\leftarrow$${\it Dec}_{1}^F(A, \{(j, y_i^j)\}_{j\in A})$ ;
     \STATE $\beta_i\leftarrow$The reconstruction phase of the SS scheme,
on input $\{s_i^j\}_{j\in A}$ ;
     \ENDFOR
     \STATE Return $g=\sum_{i=1}^{\ell}\beta_i g_i$.
    \end{algorithmic}
  \end{algorithm}
%\end{minipage}
%\end{tabular}
%\end{tabular}
\end{figure*}

The correctness of $({\it Gen}_{\le\ell}^F, {\it Eval}_{\le\ell}^F, 
{\it Dec}_{\le\ell}^F)$ just comes from the correctness of each FSS scheme
$({\it Gen}^F_{1}, {\it Eval}^F_{1}, {\it Dec}^F_{1})$
for the basis function $\chi_{a_i}(x)$ and the correctness of 
each SS scheme for the coefficients.
But some care must be done.
From the assumption, $f\in {\cal F}_{{\cal B}_F,\ell}$ has
$\ell$ terms at most. If we represent $f$ as a linear combination
of exactly $\ell$ terms, some coefficients for basis functions must be
zero. Since the $0$-function $\chi_0(x)=(\omega_q)^{0\cdot x}=1$
which maps any element $x\in \mathbb{F}_q$ to 1 
can be partitioned into several functions as the ordinary basis functions 
can be, we can apply 
$({\it Gen}^F_{\le\ell}, {\it Eval}^F_{\le\ell}, {\it Dec}^F_{\le\ell})$ 
as well.

The security
of $({\it Gen}^F_{\le\ell}, {\it Eval}^F_{\le\ell}, {\it Dec}^F_{\le\ell})$ 
can be discussed as follows.
Without of loss of generality, 
we assume that all parties in a forbidden set $B$
(where $|B|=m$)
get $((\vec{k}_1,\vec{s}_1),\ldots,(\vec{k}_m,\vec{s}_m))$.
For any $i$ with $1\le i \le \ell$,
the $m$-tuples of
the $i$-th elements of $\vec{k}_1, \ldots, \vec{k}_m$ are identical
whatever the basis function for the $i$-th term of the target function is, 
because the advantage of any adversary
against $({\it Gen}^F_{1}, {\it Eval}^F_{1}, {\it Dec}^F_{1})$ is 0
as discussed in Section 4.1.
Moreover, for any $i$ with $1\le i \le \ell$,
the $m$-tuples of 
the $i$-th elements of $\vec{s}_1, \ldots, \vec{s}_m$ are identical
whatever the coefficient for the $i$-th term of the target function is,
because of the perfect security of the underlying SS scheme with the
same access structure. Furthermore, 
the outputs of several executions of ${\it Gen}^F_1$ 
(even for the same target basis function) are independent
because each ${\it Gen}^F_1$ uses a fresh randomness.
Thus, the information that all the parties in $B$ can get is
always the same regardless of the target function 
$f\in {\cal F}_{{\cal B}_F,\ell}$.
This guarantees the security of
$({\it Gen}^F{\le\ell}, {\it Eval}^F_{\le\ell}, {\it Dec}^F_{\le\ell})$.
%is also straightforward. Since basis functions are 
%linearly independent orthogonal vectors,
%any information on some basis function $\chi_{a_i}$ is independent to
%the information of the other basis function $\chi_{a_j}$ where $i\ne j$.
%Thus, a simple reduction to
%$({\it Gen}^F_{1}, {\it Eval}^F_{1}, {\it Dec}^F_{1})$
%and the SS scheme
%guarantees 

\vskip 5mm

\noindent
{\it Remark.}
If we do not care about the leakage of the number of terms
with non-zero coefficients for $f$, we can omit the partitioning of 
zero-functions, which increases the efficiency of the scheme.

%the discrete function $f$ which is described by the composition 
%of basis function $h$ is hidden because $h$ is hidden.

\section{Conclusion}
%In this paper, the FSS scheme for point functions, interval functions and partial pattern matching function was proposed.
%But, the target of those were some limited function class.
By observing that Fourier-based FSS schemes by Ohsawa et al.\cite{OKK17}
are compatible with linear SS schemes,
we have provided Fourier-based FSS schemes with general access structure,
which affirmatively answers the question raised in \cite{OKK17}.

\section*{Acknowledgement}
TK is supported in part by JSPS Grant-in-Aids for Scientific
Research (A) JP16H01705 and for Scientific Research (B) JP17H01695.

\end{document}